\documentclass[a4paper, 11pt]{article}

\usepackage{graphicx,fancyhdr,amssymb}
\usepackage{amsmath,cite}
\usepackage{xcolor,booktabs}
\usepackage{soul}
\usepackage{url}
\usepackage{caption,authblk,cmbright}

\usepackage{hyperref}
\hypersetup{
	unicode=false,          
	pdftitle={Negative frequencies in different analogue Hawking radiation systems},    
	pdfauthor={Raul Aguero-Santacruz},     
	pdfsubject={Analogue Gravity},   
	pdfcreator={Raul Aguero-Santacruz},   
	pdfproducer={Raul Aguero-Santacruz}, 
	pdfkeywords={analogue gravity, Hawking radiation, optical fibres}, 
	pdfnewwindow=true,      
	colorlinks=true,       	
	linkcolor=magenta,      
	citecolor=blue, 			
	filecolor=red,      		
	urlcolor=red           	
}

\begin{document}

\lhead{Negative frequencies in analogue Hawking radiation systems}
\rhead{Aguero-Santacruz, Bermudez}

\title{Negative frequencies and negative norms in\\ analogue Hawking radiation systems}

\author{Raul Aguero-Santacruz\footnote{{\it email:} raul.aguero@cinvestav.mx} \ and David Bermudez\footnote{{\it email:} david.bermudez@cinvestav.mx}}
\affil{\textit{Department of Physics, Cinvestav, A.P. 14-740, 07000 Ciudad de M\'exico, Mexico}}

\vspace{10pt}

\date{}
\maketitle

\begin{abstract}
In this work, we study the core concepts of Hawking radiation in the astrophysical and analogue systems. We focus on the definitions of negative frequencies and negative norms: their relationship and their role in the particle creation process of the Hawking effect. We characterize the dispersion relation by the signs of the frequency and the norm. We conclude that the most natural frame for studying the Hawking effect is in the frame in which the horizon is static, where the sign of the norm can be made equal to the sign of the Doppler-shifted frequency in that frame. We use as examples the four most successful experimental analogue systems: water waves, Bose-Einstein condensates, polaritons fluids, and optical fibers.\\

\end{abstract}

\section{Introduction}

The process of emission of Hawking radiation in astrophysical black holes \cite{Hawking1974} remains one of the most interesting topics in physics since it bridges the usually separated fields of general relativity, quantum mechanics, and thermodynamics \cite{Bekenstein1973}. The theory developed by S. Hawking uses a semiclassical approach to study the evolution of quantum fields propagating on a classical background spacetime \cite{Hawking1975}. Black holes should then radiate via the Hawking process, where a pair of particles is produced near the event horizon, one of which can escape to infinity and be detected as radiation: This is the Hawking radiation. The second particle is trapped inside the event horizon, it is said to have negative frequency and is the Hawking partner. Hawking studied the conversion between ingoing and outgoing modes by tracing back in time an outgoing mode escaping from the horizon. He realized that this mode was created by the mixing of positive- and negative-frequency modes.

Since its proposal, the Hawking effect has influenced the development of several fields of physics, including astrophysics \cite{Abedi2023}, quantum field theory in curved spacetimes, string theory \cite{Almheiri2021}, cosmology \cite{Leonhardt2019}, optics \cite{Aguero2020}, and possibly others. However, its major drawback lies in the challenges it presents to actually measure its effects: Its direct astrophysical detection is not considered possible with current or foreseeable technology \cite{Unruh2014}. This situation changed with the proposal of W. Unruh of analogue systems, where a process similar to Hawking radiation appears \cite{Unruh1981}, raising the possibility to study this effect in the laboratory. In the decade of the 2000's, scientists worked to make this possibility a reality. There are many different physical systems where analogue Hawking radiation can be theorized and experimentally measured for either a classical or a quantum field. These systems include: water waves \cite{Rousseaux2008,Rousseaux2010}, superfluids \cite{Jacobson1998,Novello2002}, BECs \cite{Munoz2019,Bermudez2019}, optical pulses \cite{Philbin2008,Bermudez2016pra,Drori2019,Aguero2020}, and polaritons \cite{Jacquet2022,Jacquet2023}. So even more fields are impacted by Hawking radiation.

In recent years, a different approach to studying Hawking radiation has emerged in the form of analogue Hawking radiation via a variety of systems, each with a rich diversity of dynamics, but with one key element in common: they all have the same kinematics as radiation around astrophysical black holes. In each analogue system, we can identify a horizon that can also emit an analogue of Hawking radiation. An important remark is the possibility to bring these systems into the laboratory, where this effect has already been successfully measured in several systems \cite{Philbin2008,Steinhauer2018,Drori2019,Munoz2019}.

Such analogue systems share a number of common features: There is always an interaction between a classical background and a classical or quantum fluctuation. This interaction replaces gravity in the original system, arises from an interaction with a moving medium, and can be modeled as an effective curvature. The curvature includes a region that blocks ingoing or outgoing modes and resembles an event horizon. This region consists of a transonic flow, where the moving medium goes from subsonic to supersonic. When an ingoing fluctuation mode interacts with this horizon, it scatters into two new outgoing modes: the Hawking and the partner modes. This is the main feature for which we can consider them as analogue systems producing analogue Hawking radiation.

These analogue systems do not exactly reproduce what happens in astrophysical black holes. Understandably, the dynamics are different for each system. However, Hawking radiation is a phenomenon that depends only on the kinematics of the system, and these conditions can be replicated in different analogue systems \cite{Visser2002,Visser2003}.

There are only two outgoing modes in the astrophysical Hawking process due to the linear dispersion relation that determines the number of modes. However, it is possible to have more than two outgoing modes in analogue systems due to a non-trivial dispersion relation. It also causes the spectrum to lose its thermality through the appearance of a gray body factor \cite{Linder2016}. At the same time, the introduction of dispersion solves a theoretical problem in Hawking's original derivation of the apparent need of ultrahigh frequency modes for the radiation to be emitted: the Transplanckian problem \cite{Jacobson1991,Visser2003}.

Another advantage of analogue systems is the possibility to measure the negative-frequency Hawking partner mode, which is impossible to measure in astrophysical black holes \cite{Drori2019}. This possibility has drawn attention to the concept of negative frequencies and has introduced a number of experimental and conceptual challenges to understand them \cite{Rousseaux2008,Rubino2012prl,Biancalana2012,Drori2019,Aguero2023}. The importance of negative frequencies and their distinction from positive frequencies is usually discussed in the context of quantum field theory, where negative and positive frequencies are related to the annihilation and creation operators, respectively \cite{Birrell1982,Brout1995,Mukhanov2007}. For classical systems, its importance is equally valid and necessary for a proper explanation of processes where mixing of positive and negative frequencies is possible, including analogue Hawking radiation.

As is common in QFTCS, the scalar product can be a negative quantity and it defines a pseudonorm instead of a norm \cite{Greiner2000} e.g., in the Klein-Gordon equation. As a consequence, we can have modes with either a positive or a negative norm. It is usual to partition the modes into two sets: a set with positive norm and positive frequency, and a set with negative norm and negative frequency. As we will see, in practice the signs of the norm and frequency do not necessarily match.

The dispersion relation is also split into positive and negative frequency branches. It is usually convenient to express the modes of the fluctuation field in terms of these branches. We find that it is possible to match the sign of the norm to the sign of the frequency for all modes. This is not possible for any frequency, but only for the frequency defined in the reference frame comoving with the horizon. Interestingly, the negative-norm modes (negative comoving frequency modes) appear only when the fluctuation modes move from the subsonic to the supersonic region, as we will see by examining the graphical solutions in the supersonic dispersion relation.

In this work, we review Hawking radiation in astrophysics and analogue systems in Section 2. We revisit the concept of negative frequencies and how they are understood by the Doppler effect in Section 3. We discuss the role of the scalar product and the pseudo-norm and how they are related to the branches of the dispersion relation. We note the importance of the negative-frequency modes for the Hawking radiation in Section 4, where we obtain that the sign of the norm can be aligned with the sign of the frequency defined in the reference frame of the horizon. In Section 5 we briefly describe these concepts in four analogue systems: gravity waves in water tanks, fluctuations in Bose-Einstein condensates, light polaritons in a polariton fluid, and light pulses in an optical fiber.

\section{Hawking radiation in astrophysics and analogue systems}

In 1974, S. Hawking proposed a novel mechanism by which black holes formed by the gravitational collapse of a massive body, such as a star, would emit thermal radiation \cite{Hawking1974}. The energy of this radiation comes from the evaporation of the black hole, since in the overall picture the emitted radiation is accompanied by a loss of mass from the black hole. Such an important conclusion can be analyzed by studying Hawking's approach, where a pair of modes of opposite frequency appear. The positive-frequency mode with an escaping trajectory is what we interpret as Hawking radiation, while the negative-frequency mode with a falling trajectory into the black hole is called the partner and is what reduces the black hole mass. The creation of these pairs of quanta arises from the behavior of quantum fields in curved spacetimes, where the notion of vacuum depends on the observer \cite{Birrell1982,Mukhanov2007}. Therefore, two different sets of bases can be used to describe a quantum field $\hat{\phi}$ in two quasi-static and asymptotic regions of spacetime (i.e., before and after the collapse of the star) and can be written as
\begin{align}
 \hat{\phi}(x,t) = \int^\infty_0 \left( \hat{a}_\omega \phi^{\text{in}}_\omega (x,t) +\text{H.c.} \right) d\omega =  \int^\infty_0 \left( \hat{b}_\omega \phi^{\text{out}}_\omega(x,t) + \text{H.c.} \right) d\omega
\end{align}
where $\hat{a}$ and $\hat{b}$ are annihilation operators, each for a different basis of modes $\phi^{\text{in}}$ and $\phi^{\text{out}}$ and H.c. is the Hermitian conjugate. The expansion in one set of modes written as a linear combination of modes from the other set is known as a Bogoliubov transformation, i.e.,
\begin{align}\label{Bogoliubov_tr}
 \hat{\phi}^{\text{in}}_\omega (x,t) = \alpha\hat{\phi}^{\text{out}}_\omega(x,t) + \beta\hat{\phi}^{\text{out}*}_\omega(x,t)
\end{align}
If we use only positive-frequency modes from one set (i.e., the non-H.c. modes) to construct positive-frequency modes from the other set, then $\beta=0$ in the Bogoliubov transformation and the annihilation operators from one set are a linear combination of only annihilation operators from the other set. In this case, there is an agreement in the definition of the vacuum state for both sets. On the other hand, when $\beta\neq 0$, positive-frequency modes of one set are written as a linear combination of positive- and negative-frequency modes of the other set, and the definition of the vacuum state is different for each set.

The scattering of in-modes in the vacuum state at the event horizon leads to the spontaneous creation of particles in the out-modes. The calculation of the expectation value for an out-fluctuation in the in-vacuum, i.e., $\langle 0_{\text{in}}|b^+_\omega b_\omega|0_{\text{in}} \rangle$, leads to a spectrum of a bosonic thermal distribution with the profile of a perfect black body with a temperature given by:
\begin{align}
 T =\frac{\hbar \sigma}{2\pi}= \frac{\hbar c^3}{8 \pi G M},
\end{align}
where $\sigma=c^3/4GM$ is the surface gravity and M is the black hole mass. Hawking radiation is a fascinating theoretical construct that bridges the gap between general relativity, quantum mechanics, and thermodynamics in the context of astrophysical black holes. It has profound implications for the ultimate fate of these enigmatic entities and makes them an attractive research topic in modern theoretical physics \cite{Almeida2022analogue}. Even with technological advances and improvements in observational techniques, detecting Hawking radiation from astrophysical black holes remains an enormous challenge. The radiation produced is extremely faint, making it nearly impossible to observe directly with current and foreseeable technology (but see Ref. \cite{Abedi2023}).

The first idea for an analogue system in which Hawking radiation could be reproduced came from W. Unruh in 1981 \cite{Unruh1981} who, while teaching a course in fluid dynamics, came to the conclusion that the equations of motion for waves in fluid mechanics are similar to those for massless scalar fields in general relativity \footnote{Confirmed by direct communication with the author.}. Thus, a fluctuation field propagating in a flowing medium can be seen as propagating in curved spacetime. Within this moving medium with a flowing velocity profile $U(x)$ we identify a region where its velocity is equal to the velocity of the propagating perturbation $c$, so that space is divided into a subsonic $U<c$ and a supersonic $U>c$ region. This behavior is similar to the conditions found in an astrophysical event horizon. The transonic region $U=c$ defines the horizon.

One of the most distinctive aspects of analogue Hawking radiation is the presence of dispersion in the system, whereas for astrophysical Hawking radiation we assume a linear dispersion relation considering that the speed of light is constant at all frequencies. This absence of dispersion mechanisms leads to the previously discussed Transplanckian problem for astrophysical Hawking radiation \cite{Visser2003,Barbado2011}. On the other hand, by including a dispersion model, the analogue Hawking radiation solves the Transplanckian problem, but the price paid is the loss of thermality due to the appearance of a gray body factor \cite{Linder2016}.

The mathematical analysis to derive Hawking radiation in analogue systems can be done in exactly the same way as Hawking did for astrophysical black holes when we neglect higher-order dispersion effects. We assume that the flow is transonic and the location of the horizon is $x=x_h$. Obviously, the derivative of the flow velocity profile is non-zero at this point. We do a linear approximation $U(x) \approx \alpha x + c$ and it can be proved that pair production takes place and the outgoing modes will have a black body radiation profile with temperature
\begin{align}
 T =\frac{\hbar \alpha}{2\pi}, \quad \alpha = \frac{dU(x)}{dx}|_{x=x_h},
\end{align}
where $\alpha$ is the generalized surface gravity, which depends on the velocity profile of the moving medium.

\section{Negative frequencies}
We have found that in general it is better to describe the analogue event horizon in a reference frame where it is static, since experiments for each analogue system can be quite different from each other. This frame can be either the laboratory frame or a frame moving with the effective mechanism that generates the horizon, called the comoving frame. It is important to keep track of how the mode frequencies of the fluctuation field change between these two frames. The Doppler effect relates the observed frequency $\omega_o$ in one frame to the emitted frequency $\omega_e$ in the other frame
\begin{align}\label{doppler}
 \omega_o = \omega_e \left( \frac{c \pm V_o}{c \mp V_e} \right),
\end{align}
where $c$ is the velocity of the fluctuation, $V_o$ is the velocity of the observer, $V_e$ is the velocity of the emitter, the sign in the numerator is positive if the observer is moving toward the emitter, and negative if the observer is moving away from the emitter source. The sign in the denominator does the opposite.

The Lagrangian for dispersive systems is modeled with the inclusion of higher-order derivatives $L=L(\phi, \partial_t \phi, \partial_x \phi, ...)$, where $\phi$ is the fluctuation field. Using the least action principle, we can obtain the Euler-Lagrange equation, which gives the wave equation containing the dynamics of the system. The resulting dynamical equation contains the information of the dispersion relation in the reciprocal space, which can be written as
\begin{align}\label{funF}
 \omega'^2= (\omega - Uk)^2 = F(k)^2,
\end{align}
where $\omega'$ is the comoving frequency obtained by a Doppler shift from the laboratory frequency $\omega$ and known as the free-fall frequency $F(k)$. In the dispersionless case, the dispersion takes a linear form $F(k) =\pm ck$ and for dispersive cases it takes more complex functional forms. In the latter case, the dispersion relation can be classified as subsonic or supersonic, depending on the value of the phase velocity $|\omega/k|$ with respect to the value $c$ of the dispersionless case. As we will see, the transition from one region to the other is crucial for the appearance of positive- and negative-norm solutions.

If the Lagrangian is invariant to time translations, we have solutions called stationary modes, which have the form $\phi(x,t) = e^{-i\omega t} \phi_\omega (x)$. A complete solution involves the complex conjugate of the stationary modes, so it is common to express the field as an expansion of modes with positive and negative frequencies $\omega$, i.e.,
\begin{align}
 \phi(x,t) = a_\omega e^{ikx-i\omega t} +a^*_\omega e^{-ikx+i\omega t},
\end{align}
where the first term represents the positive-frequency modes and the second term represents the negative-frequencies modes.

\section{Negative norms}
As we discussed earlier, Hawking radiation is not a full quantum gravity effect, but rather a consequence of analyzing a quantum fluctuation field in a classical curved spacetime background, which is known as semiclassical gravity. Even more remarkable is the fact that in analogue systems the curved spacetime results from the consideration of an effective theory describing a moving medium in which a quantum or even a classical field fluctuation propagates. In the simplest case, such a propagating field is a massless scalar field that can be decomposed into appropriate modes. These modes can be characterized by properties of their wave nature, namely their wavenumbers or frequencies.

Therefore, the general solution is a linear combination of all possible solutions over a given basis. This linear expansion must include the space of complex conjugate solutions. The distinction between ordinary solutions and their complex conjugates gives us a natural separation for positive- and negative-frequency solutions. On the other hand, the norm of a quantum field can be computed by the definition of the scalar product
\begin{align}\label{scalar_product}
 (\phi_1,\phi_2 ) = i \int^\infty_{-\infty} dx (\phi_1^* \pi_2 - \phi_2 \pi_1^*),
\end{align}
where $\phi$ is the massless field and $\pi$ is its canonical momentum defined as
\begin{align}
 \pi = \frac{\partial L }{\partial (\partial_t \phi^*)},
\end{align}
and $L$ is the Lagrangian. Unlike traditional scalar products, Eq. \eqref{scalar_product} can be either positive or negative, and is therefore known as a pseudo-norm \cite{Greiner2000}. However, the sign of the norm and the sign of the frequency are not necessarily the same for a given solution. When we consider all possible modes that can occur given an input state, we look for modes that are solutions of the dispersion relation given the value of a conserved Killing vector, usually the laboratory frequency \cite{Robertson2012jpb}. For the special case of the fiber-optical analogue, the conserved quantity is the comoving frequency. Within these possible modes we find solutions that extend into the range of negative frequencies for the dispersion relation, and as such are part of the basis of complex conjugate solutions. As a side note, when talking about real fields, such as the electric field, its complete decomposition into complex modes and their conjugates must simultaneously include positive and negative frequencies spanned over a basis. Both sets are necessary to recover the real solution \cite{Amiranashvili2022,Aguero2023}.

When we expand the modes to form a general solution, we can choose to do so in the basis of plane waves that span both dispersion branches in Eq. \eqref{funF}. This means that we take $\omega = Uk\pm|F(k)|$ and the solution takes the form:
\begin{align}
 \phi(x,t) = \frac{1}{2\pi} \int^\infty_{-\infty} \left[ A^{+}(k) e^{ikx - i (Uk+|F(k)|)t}  + A^{-}(k) e^{ikx - i (Uk-|F(k)|)t} \right] dk.
\end{align}
From this expansion we can confirm that the exponential terms in the second sum are the complex conjugates of the exponentials in the first sum, so we have a complete set. If we use the scalar product of Eq. \eqref{scalar_product} to the previous equation, we find that its norm is
\begin{align}\label{norm}
 (\phi, \phi) = \frac{1}{\pi} \int^\infty_{-\infty} |F(k)| \left[|A^{+}|^2 - |A^{-}|^2 \right] dk.
\end{align}
If the Lagrangian for the system under analysis is invariant under global phase transformations, then the scalar product of the field is a conserved quantity. By considering a positive-norm mode as an input to the system, part of the field can be transformed into a negative-norm mode, accompanied by an increment of positive-norm modes, in such a way that the total norm is conserved. If the input mode is the vacuum state, a pair of modes with opposite norm and frequency can be created, and we have the so-called spontaneous creation of a pair of particles. The Hawking effect is an example of such an effect.

The branches in Eq. \eqref{norm} are decoupled and the sign of a solution is equal to the sign of the comoving or free-fall frequency. In fact, the modes in the positive dispersion branch with positive norm have a positive comoving frequency. This construction of the space of solutions carries with it the importance of the negative-frequencies as a core element in the description of the analogue Hawking radiation.

As we have described, the dynamics of all analogues of Hawking radiation are described by some wave equation, most of them second-order equations after some approximations. These in turn are translated into second-order algebraic equations from which we derive the dispersion relation. In fact, there are two possible solutions or branches of the dispersion relation: one for positive frequencies and another for negative frequencies. We will now discuss the dispersion relation for the four most relevant experimental systems for analogue gravity. We will describe the dispersion relations for each system in Eqs. \eqref{dispWater}, \eqref{dispBEC}, \eqref{dispPolariton}, and \eqref{dispOptics}, and their respective plots in the laboratory and in the comoving reference frame for the supersonic case in Figs. \ref{figDispWater}, \ref{figDispBEC}, \ref{figDispPolariton}, and \ref{figDispLight}.

The four most studied analogue systems are gravity waves in water tanks \cite{Rousseaux2008,Rousseaux2010}, phonons in Bose-Einstein condensates \cite{Macher2009,Recati2009,Steinhauer2018,Munoz2019}, light pulses in optical fibers \cite{Bermudez2016pra,Drori2019}, and, more recently, light polaritons in a microcavity \cite{Jacquet2022,Jacquet2023}. We briefly describe them in the following subsections, including the definition of negative frequencies and negative norm solutions.

\section{Analogue gravity systems}

\subsection{Water waves}
\begin{figure}
	\centering
	\includegraphics[width=0.49\linewidth]{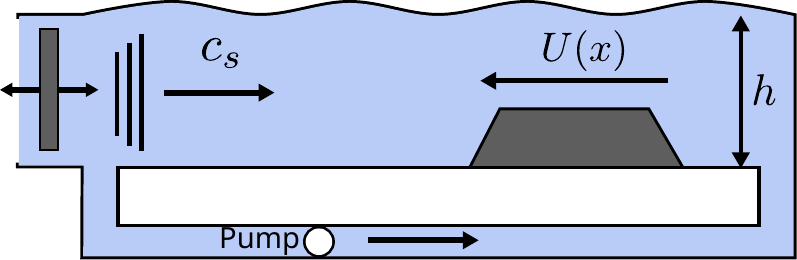}
	\caption{Diagram of an experiment to measure analogue Hawking radiation in water waves \cite{Rousseaux2008}. The analogue event horizon is stationary in the laboratory reference frame.}
	\label{figWater}
\end{figure}

In the context of water waves, the classical phenomenon of analog Hawking radiation can be observed in a flowing fluid with a velocity profile varying in space $U(x)$, as in a water channel, where we consider only one effective spatial dimension. This setup can be obtained in a long water tank with varying depth. The forces acting on the fluid are such that the horizontal flow remains stationary. We consider an irrotational, inviscid, and incompressible fluid \cite{Rousseaux2008,Weinfurtner2011}. The analogue curved spacetime is obtained by varying the depth of the bottom of the tank. This variation produces the velocity gradient flow and we can identify a horizon that is stationary in the laboratory frame. Fluctuation waves are generated by an external method (usually an oscillating paddle) and propagate against the fluid flow up to the point where their velocity $c$ equals the velocity of the flow $U$: This point is called the horizon. In Fig. \ref{figWater} we show a schematic of the water wave analogue in the water tank configuration.

The dynamics of surface waves can be described by the Klein-Fock-Gordon equation for the fluctuation $\phi$
\begin{align}\label{propWater}
 (\partial_t + \partial_x U) (\partial_t + U \partial_x) \phi = V_{\text{ext}}(-i\partial_x) \phi,
\end{align}
where $V_{\text{ext}}(i\partial_x)$ represents a generalized potential that, for the particular water tank configuration described above, can be written as
\begin{align}\label{Vext}
V_{\text{ext}}(-i\partial_x) =i \left( g\partial_x - \frac{\gamma}{\rho} \partial_x^3 \right) \tanh{(-ih\partial_x) }.
\end{align}
The dispersion relation for this system is
\begin{align}\label{dispWater}
 (\omega - U k)^2 = \left( gk + \frac{\gamma}{\rho} k^3 \right) \tanh{(kh)},
\end{align}
where $g$ is the free fall gravity acceleration, $\rho$ is the density of the fluid, $\gamma$ is the surface tension, and $h$ is the depth of the fluid in the water tank.

\begin{figure}
	\includegraphics[width=0.50\linewidth]{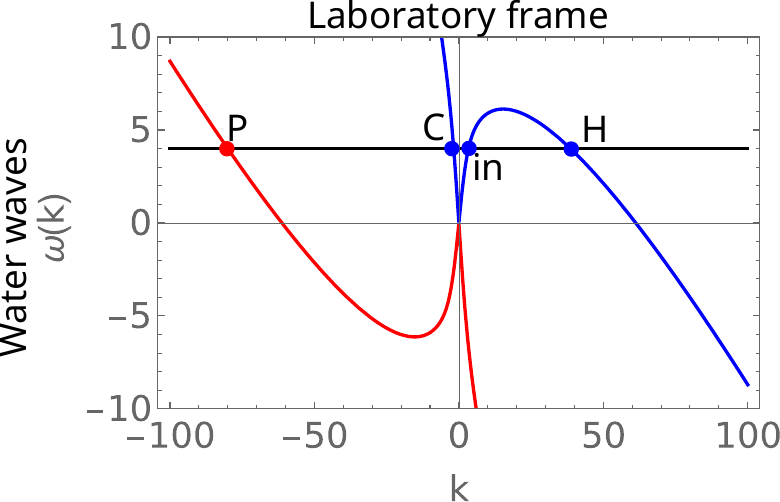}
	\includegraphics[width=0.48\linewidth]{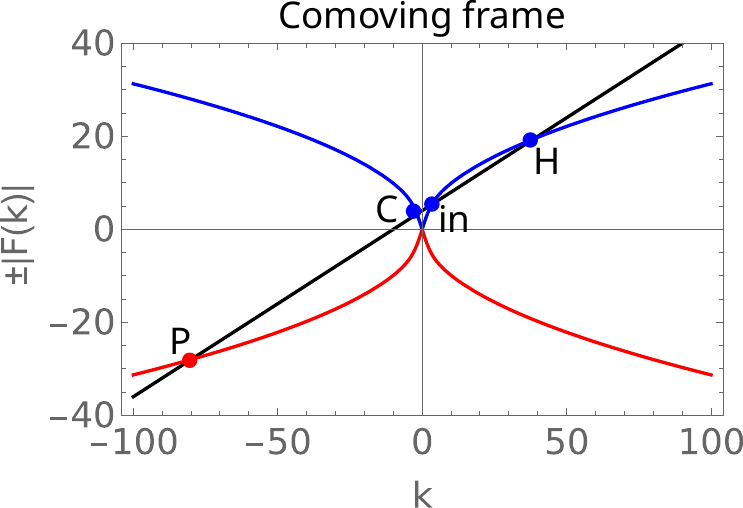}
	\caption{Left: Dispersion relation in the laboratory frame. Right: Dispersion relation in the comoving frame for the water wave analogue.}
	\label{figDispWater}
\end{figure}

Now let us define the laboratory and comoving frequencies using the Doppler effect described by Eq. \eqref{doppler}. In this case the observer is fixed in the  laboratory, i.e. $V_o = 0$, the emitter is the fluid moving with the waves at velocity $V_e=U$, the laboratory frequency is $\omega_0 = \omega$, and the comoving frequency is $\omega' =\omega_e$. For this case, the Doppler formula is then
\begin{align}
 \omega' = \omega \frac{c-U}{c}.
\end{align}
In the laboratory frame, we consider the detector's point of view to define positive frequencies, i.e. since a frequency detector only detects the beating of waves, we usually assign a positive frequency to it. In this case, all frequencies measured in the laboratory are positive quantities, but the difference $c-U$ is not always a positive quantity, so the value of $\omega'$ can have a different sign depending on the situation. There are three possibilities: (i) in the subsonic region $U<c$ then $\omega' >0$, (ii) in the transonic $U=c$ then $\omega'=0$, and (iii) in the supersonic $U>c$ then $\omega'<0$. The supersonic flow condition is accompanied by negative comoving frequencies. This is the case shown in Fig. \ref{figDispWater} where we plot the dispersion relation Eq. \eqref{dispWater} in the laboratory and comoving frames. There is a transition condition $U=c$ and $\omega'=0$ where negative comoving frequencies appear: This is the so-called horizon frequency. In dispersive systems it is more useful to define the horizon as a frequency rather than a spatial point \cite{Bermudez2016pra}. This is the point where an ingoing mode undergoes a frequency shift and part of its norm is transformed into a pair of positive- and negative-frequency modes by norm conservation.

If we compute the norm for this system, we would find that Eq. \eqref{norm} can be expressed as
\begin{align}\label{normWater}
 (\phi, \phi) = \frac{1}{\pi} \int^\infty_{-\infty} \left| \left( g\partial_x - \frac{\gamma}{\rho} \partial_x^3 \right) \tanh{(-ih\partial_x) } \right| \left[|A^{+}|^2 - |A^{-}|^2 \right] dk.
\end{align}

In Fig. \ref{figDispWater} the positive and negative branches of the supersonic dispersion relation are shown in blue and red, respectively and the black line represents a given conserved frequency. The possible solutions are the modes resulting from the intersection of the black line with one of the two branches. The Hawking effect consists of the emission of two signals of opposite-sign norm, one solution in the positive branch and one solution in the negative branch. In both the laboratory and the comoving system, there are three solutions in the positive branch and one in the negative branch. The input fluctuation (in) is the solution near the origin with $k>0$ and positive norm $+|F(k)|$, the Hawking mode (H) has larger $k\gg 0$. The partner mode (P) has $k\ll0$ and negative norm $-|F(k)|$. The fourth mode is counterpropagating (C) and does not affect the Hawking process.

Before the fluctuation reaches the horizon, the dispersion relation is that of the subsonic case, where there are only two positive-norm solutions, excluding the Hawking modes. Since there are no modes in the negative branch, $A^- =0$ in the norm Eq. \eqref{normWater}. It is only when the fluctuation passes the horizon condition and enters the supersonic condition $U<c$ that the negative-norm modes become real solutions and contribute to the norm with a factor $-|F(k)| |A^-|^2$. The supersonic condition also defines the negative frequencies for the frame where the detector is not static due to the Doppler effect. Thus, this condition not only defines an exotic regime, but also describes the existence of negative-norm modes and their association with negative-frequency modes.


\subsection{Bose-Einstein condensates}
Now we consider a Bose-Einstein condensate (BEC) in a quasi-one-dimensional laser trap, elongated enough such that its transverse dynamics can be considered constant, and the system is reduced to an effective one-dimensional problem. The condensate flows with a spatial variation of its velocity $U(x)$ and density profile $n(x)$ near a region where the flow is kept stationary in time by interaction with an external potential $V_{\text{ext}}(x)$ together with tuning of the interaction constant $g(x)$. Within the condensate, the speed of sound is $c(x) = \sqrt{gn(x)/m}$, where $m$ is the mass of the condensate. The desired velocity profile to obtain a black hole configuration requires the implementation of a sink of atoms, which is physically achieved by a moving laser sweeping the condensate cite{Steinhauer2016}. Therefore, we have two regions separated by this sink: a subsonic half where $U<c$ usually called the upstream ($u$) region, and a supersonic half where $U>c$, called the downstream ($d$) region. In this way it is possible to obtain an effectively curved spacetime in which fluctuations can propagate. Fluctuations in the $d$ region cannot access the $u$ region: This is the analogue of the interior of an astrophysical black hole, and the sink separating the two regions which satisfies the sonic condition $U=c$ is equivalent to the event horizon.

The condensate wave function $\hat{\psi}$ models the fluctuation and is a solution of the stationary Gross-Pitaevskii equation
\begin{align}\label{propBEC}
 \left[ -\frac{\hbar^2}{2m} \partial_x^2 + V_{\text{ext}}(x) +g(x) |\hat{\psi}|^2\right] \hat{\psi} =\mu \hat{\psi},
\end{align}
where $\mu$ is the chemical potential. Far away from the sink both regions reach an asymptotic limit where the flow and fluctuation velocities reach a static value $V_{u,d}$ and $c_{u,d}$, respectively. In these limits, the dispersion relation can be written as
\begin{align}\label{dispBEC}
F(k)^2 = (\omega - U_{u,d} k)^2  =  c^2_{u,d} k^2 \left( 1+ \frac{1}{4} (k\xi_{u,d})^2 \right),
\end{align}
where $\xi_{u,d} =\hbar /(m c_{u,d})$ is the healing length, which describes the interaction range of the condensate in the two $u,d$ regions.

\begin{figure}
	\centering
	\includegraphics[width=0.49\linewidth]{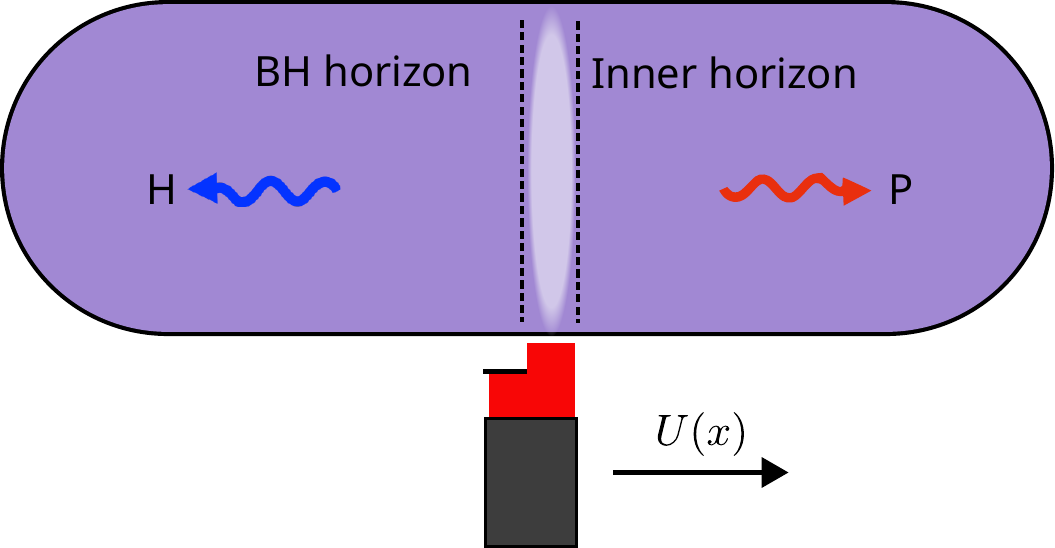}
	\caption{Analogue Hawking radiation for Bose-Einstein condensates. The analogue event horizon is stationary in the comoving reference frame.}
	\label{figBEC}
\end{figure}

In this system, the moving laser creates the necessary potential that produces the separation of the medium by means of the analogue event horizon. In the laboratory frame, the horizon moves along with the laser, but we can define a comoving frame where it is stationary. We show a schematic of the system in Fig. \ref{figBEC}. To use Eq. \eqref{doppler}, we identify the observed frequency as the comoving frequency $ \omega_o =\omega'$ and the emitted frequency as the laboratory frequency $\omega_e = \omega$. The negative of the flow velocity replaces the velocity $V_o=-U$ and the velocity of the emitter is $V_e=0$ in this reference frame. Thus, the frequencies in both reference frames are related as
\begin{align}\label{omega_comov}
 \omega = \omega' \frac{c}{c-U}
\end{align}
Since the optical detector also moves with the laser \cite{Steinhauer2016}, there is an effective change of roles, since now $\omega'>0$. The same three cases subsonic, transonic and supersonic that exist for water waves are reproduced here. The laboratory frequency $\omega$ can be positive or negative with an indefinite value at the transonic condition $U=c$, which describes the horizon. Even if the horizon is stationary in the comoving frame, the same physics applies to describe the core phenomenon.

By evaluating Eq. \eqref{norm} to find the norm we have
\begin{align}
(\hat{\psi}, \hat{\psi}) = \frac{1}{\pi} \int^\infty_{-\infty} \left| c_{u,d} k \sqrt{1+ \frac{1}{4} (k\xi_{u,d})^2 }\right| \left[|A^{+}|^2 - |A^{-}|^2 \right] dk,
\end{align}
where we can see again that the norm is determined by the free-fall frequency now given by Eq. \ref{dispBEC}.

\begin{figure}
	\includegraphics[width=0.51\linewidth]{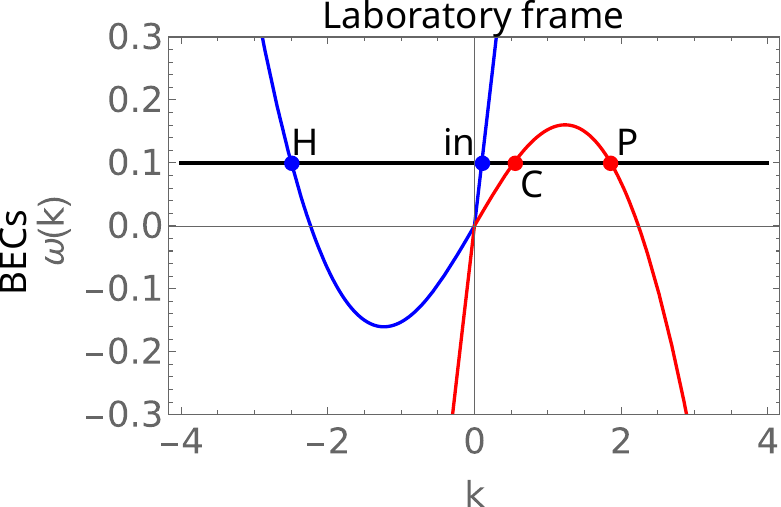}
	\includegraphics[width=0.47\linewidth]{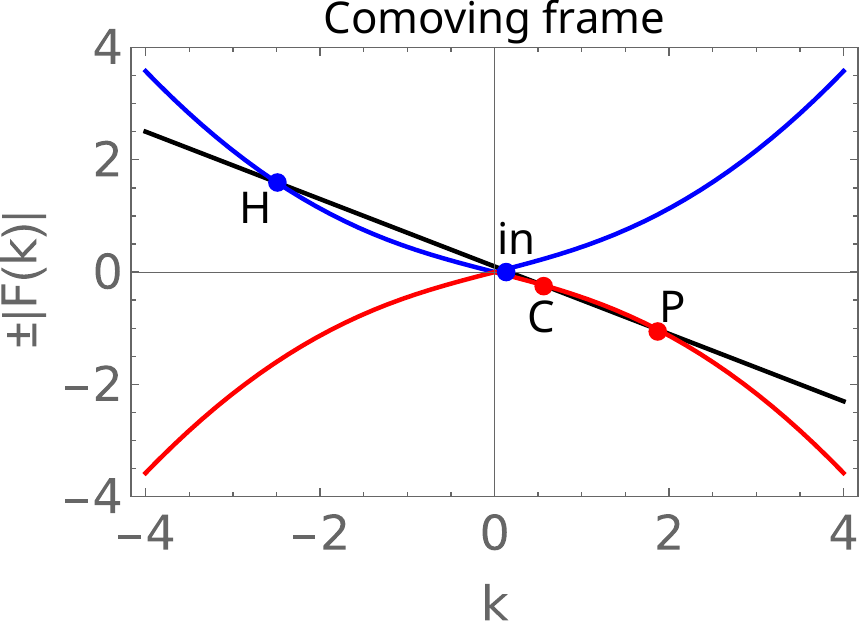}
	\caption{Left: Dispersion relation in the laboratory frame. Right: Dispersion relation in the comoving frame for the BEC case.}
	\label{figDispBEC}
\end{figure}

In Fig. \ref{figDispBEC} we plot the dispersion relation in Eq. \eqref{dispBEC} in the laboratory and comoving frames, where we have two positive-norm solutions and two negative-norm solutions. This mode structure is different from the water wave analogue.

\subsection{Polaritons fluids}
\begin{figure}
	\centering
	\includegraphics[width=0.49\linewidth]{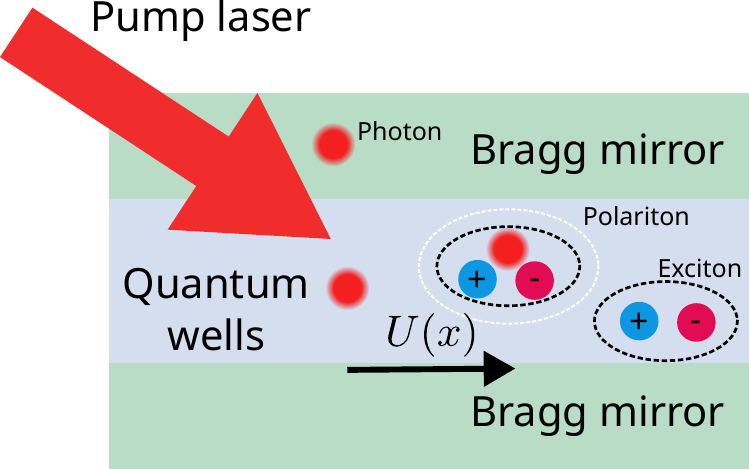}
	\caption{Analogue Hawking radiation for polariton fluids. The analogue event horizon is stationary in the comoving laboratory frame.}
	\label{figPolariton}
\end{figure}

In polariton fluids, the system consists of a semiconductor microcavity placed between two Bragg mirrors to form a thin cavity (see Fig. \ref{figPolariton}). The material inside the cavity usually consists of InGaAs quantum wells. These quantum wells confine excitons, which are quasi-particles of bound states composed of electrons and holes. These excitons interact strongly with confined photons emitted by a laser in the cavity. A combination of excitons and photons can merge to form a quasi-particle resulting from light-matter interactions within this microcavity, known as a polariton.

The Hamiltonian of this system has two eigenstates: the lower and upper polariton branches. The pump laser is chosen to be quasi-resonant with the lower branch. By introducing of a deformation within the cavity, a polariton flow can be modified to have an effectively curved spacetime for fluctuations in the polariton fluid. Then the condition to have a horizon is to have subsonic $U<c$ and supersonic $U>c$ regions separated by a transonic boundary $U=c$, which is the analogue event horizon. This system can be configured in one or two dimensions \cite{Nuyen2015acoustic,Jacquet2020}, including a rotating configuration \cite{Vocke2018rotating}.

The dynamics of a fluctuation is given by a modified Gross-Pitaevskii equation \cite{Jacquet2022}:
\begin{align}\label{propPolariton}
 i\hbar \partial_t \hat{\Psi} = \left[ -\frac{\hbar}{2m} \partial_x^2 + \hbar g + V_{\text{ext}} -i\Gamma   \right] \hat{\Psi} + F_p,
\end{align}
where $\Gamma$ is the loss rate and $F_p$ is the field of the pump laser. The dispersion relation is
\begin{align}\label{dispPolariton}
(F(k)+i\Gamma)^2 = (\omega(k) -Uk+i\Gamma)^2 = \frac{\hbar k^2}{2m} \left(\frac{\hbar k^2}{2m} +2gn  \right),
\end{align}
where $n=|\Psi|^2$. In this system, the flow is generated by the moving polaritons due to the applied pump laser. The analogue event horizon is stationary in the laboratory frame and the transonic condition $U=c$ occurs near the position of a defect introduced into the cavity, similar to previous cases. The calculated norm for this system is
\begin{align}
 (\hat{\Psi}, \hat{\Psi}) = \frac{1}{\pi} \int^\infty_{-\infty} \left| \sqrt{\frac{\hbar k^2}{2m} \left(\frac{\hbar k^2}{2m} +2gn  \right)} - i\Gamma  \right| \left[|A^{+}|^2 - |A^{-}|^2 \right] dk.
\end{align}

\begin{figure}
	\includegraphics[width=0.51\linewidth]{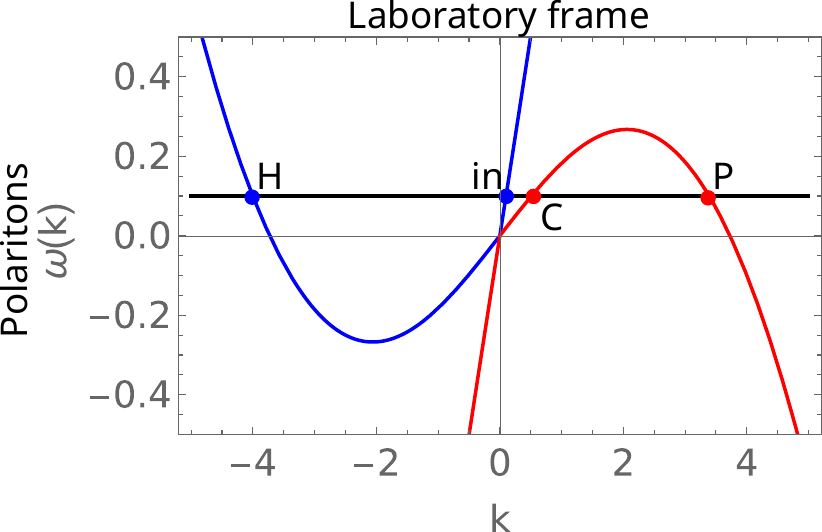}
	\includegraphics[width=0.47\linewidth]{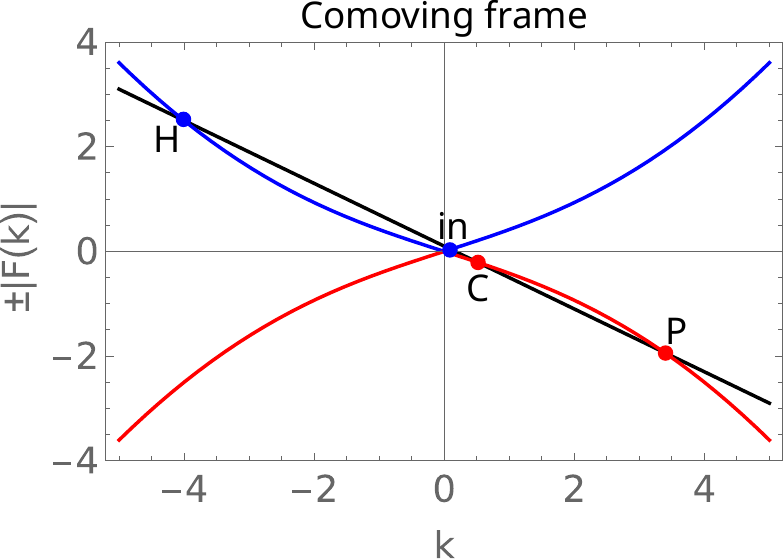}
	\caption{Left: Dispersion relation in the laboratory frame. Right: Dispersion relation in the comoving frame for the polariton case.}
	\label{figDispPolariton}
\end{figure}
In Fig. \ref{figDispPolariton} we show the dispersion relation for the polariton fluid in both the laboratory and the comoving frame. Since the functional form in Eqs. \eqref{dispBEC} and \eqref{dispPolariton} is proportional, we have a similar feature set of solutions for the BEC system, including the minus sign in the norm to have modes in and H with positive norm.

\subsection{Optical fibers}
\begin{figure}
	\centering
	\includegraphics[width=0.49\linewidth]{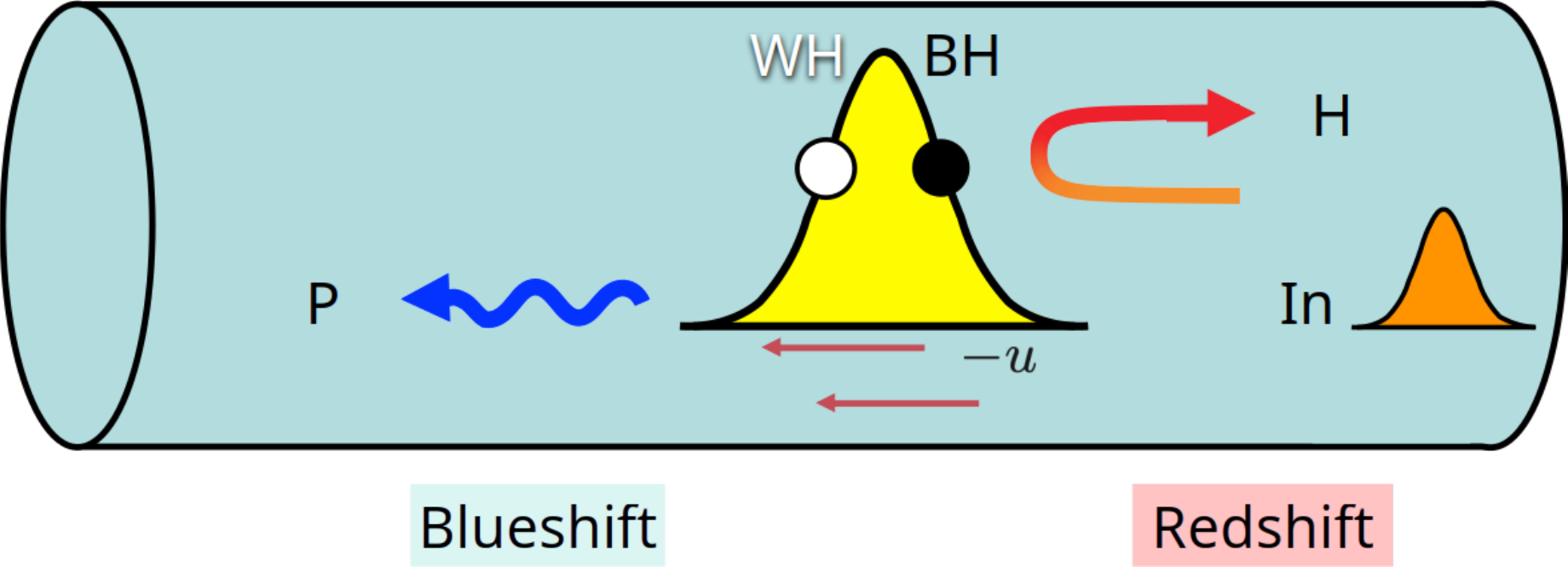}
	\caption{Diagram of the situation for analogue Hawking radiation inside an optical fiber in the comoving frame. The analogue event horizons for white hole and black holes are stationary in the comoving frame. The Hawking radiation and its partner appear when a fluctuation interacts with the pump pulse.}
	\label{figLight}
\end{figure}

In the fiber-optical analogue, the moving medium is the fiber itself, as viewed from the comoving frame of a pump pulse traveling along the fiber. It is desirable to have the shortest pump pulse possible, since it enhances the Hawking effect due to increased nonlinearity \cite{Leonhardt2009b}. As the pump pulse propagates through the fiber, there is a local change in the refractive index proportional to the intensity of the pulse $\delta n(\tau) \propto I(\tau)$, where $\tau$ is the delay time and $I$ is the intensity of the pump pulse. This phenomenon is known as the optical Kerr effect \cite{Agrawal2013}, and it is the way to obtain an effectively curved spacetime. The use of highly-nonlinear fibers known as photonic-crystal fibers (with high nonlinear coefficient $\gamma \sim 10^4 \text{W}^{-1} \text{km}^{-1}$ \cite{Russell2003photonic}) and the use of ultra short pulses (FWHM $\sim 10$ fs) are also desirable to enhance the nonlinear effects. In this way, we work in the regime of extreme nonlinear optics (XNLO) \cite{Aguero2020}, where the negative branch of the dispersion relation can be seeded to produce negative-norm solutions.
In Fig. \ref{figLight} we can see a representation of how Hawking radiation takes place inside an optical fiber when the system is viewed in the pump pulse reference frame.

Fluctuation waves traveling inside the fiber but away from the pump pulse are affected only by the refractive index of the fiber $n_0(\omega)$, which changes the speed of the waves according to their frequencies. When these fluctuation waves approach the pump pulse, the total refractive index $n(\omega,\tau) = n_0(\omega) + \delta n (\tau)$ is higher. As the fluctuation wave catches up with the pump pulse, its velocity decreases until it eventually stops at a point that is the analogue horizon. Fluctuations moving in the same direction as the effective fluid flow, i.e. the effective moving fiber in the comoving frame, are said to be copropagating, and counterpropagating when they move in the opposite direction. If the waves are counterpropagating, they reach the pump pulse through an analogue white-hole horizon or if the waves are copropagating, they reach the analogue black-hole horizon of the pump pulse. In either case, a pair of Hawking radiation signals is produced. They can be classical or quantum depending on the type of input fluctuation considered. A classical input can be another pulse, such as a soliton or a continuous wave. For the quantum effect to occur, the quantum vacuum acts as the input fluctuation. The classical case has been experimentally verified \cite{Philbin2008}, including the detection of the negative-frequency mode \cite{Drori2019}. The quantum case has not yet been experimentally verified.

The dynamics of light interacting inside an optical fiber is described by Maxwell's equations. However, the wave equation can be simplified with a minimum of approximations while still retaining validity for extreme nonlinear optics \cite{Couairon2011}. This equation is known as the unidirectional pulse propagation equation (UPPE) and is written as
\begin{align}\label{propOptics}
 i \partial_z \mathcal{E}_\omega + k(\omega) \mathcal{E}_\omega + \frac{\omega}{2cn(\omega)} \chi^3 (\mathcal{E}^2 \mathcal{E}^*+\mathcal{E}^{*2}\mathcal{E}+\mathcal{E}^3)_{ \omega +} =0,
\end{align}
where $\mathcal{E}_\omega$ is the Fourier transform of the analytic signal of the electric field $E$, $k(\omega)=\omega n(\omega)/c$ is the propagation constant or wavenumber and represents the dispersion of the system, $n(\omega)$ is the total index of refraction, and the plus subscript in the last term means positive filtering of frequencies, i.e. $\Theta(\omega)$.

In optics, the dispersion information is usually encoded in the refractive index $n(\omega)$ or the wavenumber $k(\omega)$. However, for horizon physics, when we are in the comoving frame of the pump pulse, it is more convenient to use the Doppler frequency $\omega'$ since it is (at least approximately) a conserved quantity
\begin{align}\label{dispOptics}
 \omega' (\omega) =  \gamma_L [\omega \mp U k(\omega)],
\end{align}
where $\gamma_L$ is the Lorentz factor.

In this system, the pump pulse propagating through the fiber generates the horizon, which is stationary in the comoving frame of the pulse. This is the same situation as in BECs, and the same analysis of the relationship between the different frame frequencies applies. This is $V_o = -U$, $V_e = 0$, $\omega_o =\omega'$, $\omega_e =\omega$, and therefore the Doppler shifted frequency in Eq. \eqref{doppler} takes the same form as in Eq. \eqref{omega_comov} and can be effectively reduced to Eq. \eqref{dispOptics} if we include the $\gamma$ factor in $\omega'$.

The transformation from the laboratory reference frame to the comoving frame is done by the non-boosted coordinate transformation $t =\tau-z/U$ and $\zeta = z/U$, where $\tau$ and $\zeta$ are the delay time and the propagation time, respectively.

The two coordinates $\tau$, $\zeta$ in the comoving frame have dimensions of time, $\tau$ playing the role of space and $\zeta$ that of time, and their conjugated variables are $\omega$ and $\omega'$, respectively. With the fluctuation modes now expressed in terms of the variables of the comoving frame, the two branches of the dispersion relation can be used to write a general solution to the propagation equation, which has the form
\begin{align}
 \mathcal{E} (\tau,\zeta) = \frac{1}{2\pi} \int^\infty_{-\infty} \left(A^+(\omega) e^{-i\omega \tau -i(\omega-U|k(\omega)|)\zeta} +A^-(\omega) e^{-i\omega \tau -i(\omega+U|k(\omega)|)\zeta}   \right) d\omega.
\end{align}
The norm is calculated by replacing the previous expression in Eq. \eqref{norm} and we find it to be
\begin{align}
 (\mathcal{E},\mathcal{E}) = \frac{1}{\pi} \frac{\epsilon_0 c^2}{U} \int^{\infty}_{-\infty} |k(\omega) | \left[ |A^+(\omega)|^2  -|A^- (\omega)|^2 \right] d\omega,
\end{align}
where we can see that $k(\omega)$ plays the role of the free-fall frequency and defines the sign of the norm.

\begin{figure}
	\includegraphics[width=0.51\linewidth]{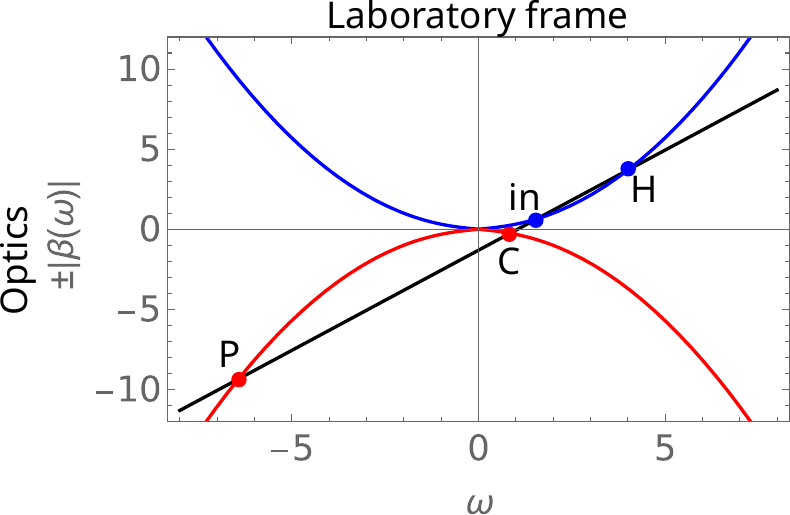}
	\includegraphics[width=0.47\linewidth]{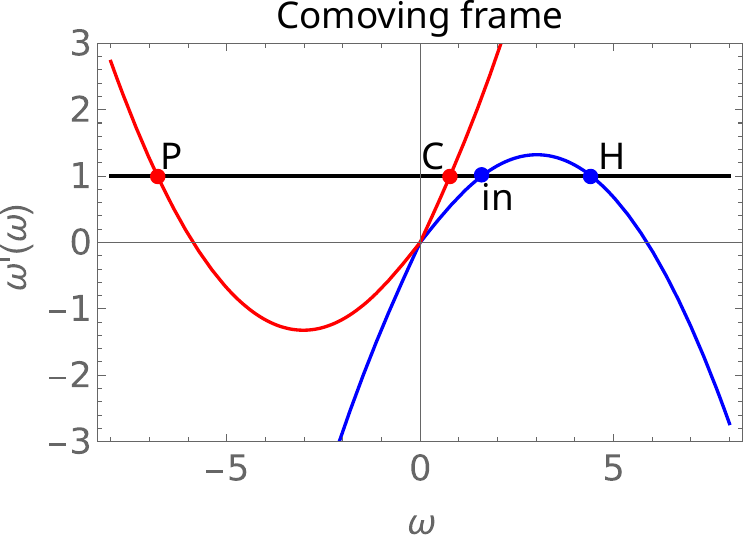}
	\caption{Left: Dispersion relation in the laboratory frame. Right: Dispersion relation in the comoving frame for the optical case.}
	\label{figDispLight}
\end{figure}

In Fig. \ref{figDispLight} we plot the dispersion Eq. \eqref{dispOptics} for the laboratory and comoving frames, where we see two positive-norm solutions and two negative-norm solutions. Also, $k(\omega)$ determines the norm, but the comoving frequency $\omega'$ is conserved. Therefore, all four systems share the same feature of presenting possible solutions with negative comoving frequencies in the superluminal region.


\section{Recent developments and conclusions}

All the systems discussed so far represent a way to study Hawking radiation in systems different from the original astrophysical one, each with its own advantages and disadvantages. In these systems there is a flowing medium over which the propagating fluctuations can be dynamically studied with a wave equation. Embedded in these equations is information about the dispersive properties of the system, and as we have seen, this property is strongly related to the mixing of positive and negative norm modes that is necessary for the Hawking effect to occur. In addition, the magnitude of the velocities involved in the horizon frame are slow.

The water waves system was the first to be theorized and the first to be taken to experimentally verified, although it is a classical effect. Even then there were measurements of the negative-frequency Hawking modes \cite{Rousseaux2008,Rousseaux2010}. The study of this system has developed gracefully thanks to the collaboration of many people \cite{Unruh1981,Rousseaux2008,Weinfurtner2011}. Recently there have been efforts to explore the quantum aspects of rotating fluids in vortex configurations \cite{Vsvanvcara2023} where the fluid temperature can be lowered and it seems possible to study quantum effects.

The BEC systems have received great attention since their inception and have been a strong candidate for experimental measurement of the spontaneous Hawking radiation. In recent years, there have been reports of such experiments \cite{Steinhauer2016,Steinhauer2018,Munoz2019,Steinhauer2022} by correlating the positive and negative Hawking modes.

The polariton system has recently undergone a strong theoretical development \cite{Jacquet2022,Jacquet2023}. There is also current ongoing experimental efforts that are expected to give results soon. Even though this system may show similarities to the BEC system, its range of possible configurations makes it an interesting choice for many more studies to come \cite{Claude2023}. Including a 2D rotating system to study rotating black holes (Kerr solution) and other phenomena such as superradiance.

The optical fiber analogue, while differing from the others in the way some of its variables are represented, such as the substitution of $k$ and $\omega$ for $\omega$ and $\omega'$, retains the same core features. There is a good theoretical understanding of the effect and the experimental verification for a classical fluctuation has been reported in the measurement of the positive and negative Hawking modes \cite{Philbin2008,Bermudez2016pra,Drori2019,Felipe2022}.

The study of negative frequencies is important for the analysis of Hawking radiation, either for astrophysical or analogue black holes. Together with the norm and the dispersion relation, we can gain very useful insight into how positive and negative are related, and why modes with a certain norm may be easier to detect than others.

The important supersonic condition $U>c$ is what gives rise to the negative norm modes, which are solutions of the negative-frequency branch in this regime, and the sign of the norm is equal to the sign of the comoving frequency. This connection is helpful because it allows us to better understand the physical nature behind spontaneous particle creation for any analogue system. We have a clear separation of positive- and negative-frequencies and positive- and negative-norm modes that are present for both classical and quantum analogue fluctuations.

As we have seen, all analogue systems can be studied under the same general framework in which we can note their similarities and differences. These properties are particularly advantageous as they stimulate a rich research and flow of ideas within the study of analogue gravity.

\section*{Acknowledgements}
We acknowledge funding by Conahcyt Mexico Ciencia de Frontera 51458-2019. RAS acknowledges funding by Conahcyt scholarship 485053.

\bibliographystyle{unsrt}

\bibliography{samplebib}

\end{document}